\documentclass[12pt,preprint]{aastex}

\slugcomment{}

\shorttitle{Planets orbiting HD 154672 and HD 205739}
\shortauthors{L\'opez-Morales et al.}

\begin{document}

\title{Two Jupiter-Mass Planets Orbiting HD 154672 and HD 205739\altaffilmark{1}}

\author{Mercedes L\'opez-Morales\altaffilmark{2,3,4}, R. Paul Butler\altaffilmark{2}, Debra A. Fischer\altaffilmark{5}, Dante Minniti\altaffilmark{6, 7}, Stephen A. Shectman\altaffilmark{8}, Genya Takeda\altaffilmark{9}, Fred C. Adams\altaffilmark{10}, Jason T. Wright\altaffilmark{11}, Pamela Arriagada\altaffilmark{6}}

\email{mercedes@dtm.ciw.edu, fischer@stars.sfsu.edu, paul@dtm.ciw.edu, dante@astro.puc.cl, shec@ociw.edu, genya@u.northwestern.edu, fca@umich.edu, parriaga@astro.puc.cl}

\altaffiltext{1}{This paper is based on data gathered with the 6.5-meter Magellan Telescopes located at Las Campanas Observatory, Chile.}
\altaffiltext{2}{Carnegie Institution of Washington, Department of Terrestrial
Magnetism, 5241 Broad Branch Rd. NW, Washington D.C., 20015, USA}
\altaffiltext{3}{Hubble Fellow}
\altaffiltext{4}{Carnegie Fellow 2004--2007}
\altaffiltext{5}{Department of Physics and Astronomy, San Francisco State University, San Francisco, CA 94132}
\altaffiltext{6}{Department of Astronomy, Pontificia Universidad Cat\'olica, Casilla 306, Santiago 22, Chile}
\altaffiltext{7}{Specola Vaticana, V-00120 Citta' del Vaticano Italy}
\altaffiltext{8}{The Observatories of the Carnegie Institution of Washington, 813 Santa Barbara St., Pasadena, CA 91101}
\altaffiltext{9}{Department of Physics and Astronomy, Northwestern University,
2145 Sheridan Road, Evanston, IL 60208}
\altaffiltext{10}{Michigan Center for Theoretical Physics, Physics Department, University of Michigan, Ann Arbor, MI 48109; Astronomy Department, University of Michigan, Ann Arbor, MI 48109}
\altaffiltext{11}{Department of Astronomy, Cornell University, 226 Space Sciences Building, Ithaca, NY 14853}

\begin{abstract}
We report the detection of the first two planets from the N2K Doppler 
planet search program at the Magellan telescopes. The first planet has 
a mass of $M \sin i = 4.96 M_{Jup}$ and is orbiting the G3 IV star HD154672 with an orbital
period of 163.9 days. The second planet is orbiting the F7 V star
HD205739 with an orbital period of 279.8 days and has a mass of $M
\sin i = 1.37 M_{Jup}$. Both planets are in eccentric orbits, with
eccentricities $e$ = 0.61 and $e$ = 0.27,
respectively. Both stars are metal rich and appear to be
chromospherically inactive, based on inspection of their Ca II H and K
lines. Finally, the best Keplerian model fit to HD205739b shows a
trend of 0.0649 m s$^{-1}$ day$^{-1}$, suggesting the presence of an additional outer
body in that system.
\end{abstract}

\keywords{Stars: Planetary Systems --- Techniques: Radial Velocities --- stars: individual (HD 154672, HD 205739)}

\section{Introduction} \label{sec:intro}
Ongoing Doppler radial velocity surveys of nearby stars have detected
over 200 extrasolar planets in the past decade \citep{Butler06}.
Those surveys focus on late F, G, K, and M dwarfs within 50 pc and
most of the planets they have found to date are more massive than Saturn, and
are presumably gas giants.
Recently, several Neptune-mass and lower mass planets have been
detected, most of them with orbital periods of a few days
\citep{Butler04, McArthur04, Santos04, Bonfils05, Bonfils07, Rivera05,
Udry06, Udry07, Endl08}

As the search for new extrasolar planets continues, Doppler surveys
now look through a broad parameter space, including long-period Jupiter
analogs, very low mass planets in short-period orbits, multiple
planetary systems, and new planets around stars with spectral types
that extend beyond the ones traditionally been searched, i.e.  K0V to
F8V. The N2K program \citep{Fischer05} is a Doppler survey with
distributed observing campaigns at the Keck, Magellan and Subaru
telescopes, and is primarily aimed at increasing the number of known
hot Jupiters.  Because of their proximity to the host stars, the
atmospheres of hot Jupiters can be as hot as 2000 K, resulting in
detectable emission at IR wavelengths.  This makes these short period
planets ideal targets for spaceborn follow up to observe exoplanet
atmospheres \citep{Harrington07}, especially when they transit their
host stars and undergo a secondary eclipse so that small differential
changes in emission from the star-planet system can be measured
\citep[e.g.][]{Charbonneau05, Knutson07}.  It is also important to
have a sample of hot Jupiters that is large enough to provide
meaningful constraints on the formation, migration, and evolution
mechanisms of these planets \citep{Ford06}.
 
The N2K program searches a fresh 
sample of the ``next 2000'' stars, not included in other current Doppler 
surveys, by extending the search radius to distances of  
100--110 parsecs from the Sun. This project is intentionally biased towards 
metal rich stars to exploit the correlation between formation 
of gas giant planets and high stellar metallicity \citep{Santos04, Fischer05b}.
The search strategy is optimized for the detection of 
Jupiter-mass planets with orbital periods shorter than 14 days 
by obtaining radial velocity measurements on three consecutive nights.
Variability in the velocity of the host star over this short timescale 
flags the star as a candidate host for a short-period planet. 
However, gas giant planets in longer orbits are also identified
and detected with additional observations. As an example, the N2K programs at
the Keck and Subaru telescopes have already detected seven planets with periods between 21 days and 3.13 years \citep{Fischer07, Robinson07}.

Here, we report the detection of the first two planets from the
Magellan N2K program. This program has been underway at the Magellan
telescopes at Las Campanas Observatory in Chile since 2004, and
includes $\sim$ 300 FGK metal rich stars at distances between 50 and
100 parsecs.  The new planets are orbiting the stars HD 154672 and HD 205739.
Both host stars showed significant radial velocity scatter in the
first year of observations and were initially flagged as hot Jupiter
candidates. However, follow-up observations over subsequent years
revealed the presence of longer period planets and a trend on the
radial velocity curve of HD 205739 that continues to increase after
over 3.5 years.

\section{Characteristics of the Host Stars} \label{sec:chars}

HD 154672 is classified as a G3IV star, with apparent magnitude V =
8.21 and color B--V = 0.71 \citep[Hipparcos Catalog;][]{ESA97}. The
Hipparcos parallax of the star is 15.2 $\pm$ 1.11 milliarcseconds,
placing it at a distance of 65.8 $\pm$ 4.8 parsecs. The distance and
the apparent magnitude of the star give an absolute visual magnitude of $M_{V}$
= 4.12. The bolometric luminosity of the star is $L_{bol}$ = 1.88
$L_{\odot}$, where we have included a bolometric correction of -0.09
derived from the empirical transformations of \citet{Vandenberg03},
using the effective temperature, surface gravity and metallicity of
the star.  Our high resolution spectroscopic analysis, described in
\citet{Valenti05}, yields $T_{eff}$ = 5714 $\pm$ 45 K, log g = 4.25
$\pm$ 0.08, $v\sin i$ = 1.0 $\pm$ 0.5 km s$^{-1}$, and [Fe/H] = +0.26
$\pm$ 0.04 for HD154672.  The radius of the star derived from the
Stefan-Boltzmann relation and the values of the luminosity and
effective temperature above is 1.39 $R_{\sun}$.  We have also derived
a stellar mass of 1.06 $M_{\sun}$, a radius of 1.27 $R_{\sun}$ and an
age of about 9.3 Gyr using the \citet{Takeda07} grid of evolutionary
models, based on the Yale Stellar Evolution Code and tuned to the
uniform spectroscopic analysis of \citet{Valenti05}. The resultant log
g is 4.26$^{+0.06}_{-0.05}$, in agreement with the results of the
spectroscopic analysis. The uncertainties in these parameters
correspond to a 95\% credibility interval using Bayesian posterior
probability distributions. The derived stellar parameters of HD154672
are summarized in the second column of Table 1.

The second star, HD 205739, is a F7V with V = 8.56 and B--V = 0.546
\citep[Hipparcos Catalog;][]{ESA97}. The Hipparcos parallax of the
star is 11.07 $\pm$ 1.12 milliarcseconds, placing it at a distance of
90.3 $\pm$ 9.1 parsecs. This sets the absolute magnitude of HD 205739
to $M_{V}$ = 3.78, and its bolometric luminosity to $L_{bol}$ = 2.3
$L_{\odot}$.  The value of $L_{bol}$, includes a bolometric correction
of -0.03 derived from the same empirical transformations of
\citet{Vandenberg03} mentioned above. Our spectroscopic analysis
yields a $T_{eff}$ = 6176 $\pm$ 44 K, log g = 4.21 $\pm$ 0.08, $v\sin
i$ = 4.5 $\pm$ 0.5 km s$^{-1}$, and [Fe/H] = +0.187 $\pm$ 0.05. Using
the effective temperature and stellar luminosity with the
Stefan-Boltzmann relation, we calculate the radius of the star to be
1.33 $R_{\sun}$. The stellar mass, radius and age derived from the
\citet{Takeda07} grid of evolutionary models are in this case 1.22
$M_{\sun}$, 1.33 $R_{\sun}$, and 2.9 Gyr, and log g is
4.29$^{+0.06}_{-0.05}$. The parameters of HD205739 are summarized in
the last column of Table 1.

Finally, the Ca II H \& K lines of HD 154672 and HD205739 (Figure 1) indicate
that their chromospheric activity is low. We can therefore reject
activity as the cause of the observed radial velocity variations of
the stars. Based on these observations, we adopt a conservative upper
limit to the expected jitter (or astrophysical noise) of the stars of about
4 m s$^{-1}$ \citep{Wright04}.

\section{Doppler Observations and Keplerian Fits}\label{sec:dopobs}

Doppler observations were carried out between mid 2004 and February
2008 at the Magellan Clay telescope using the MIKE spectrograph
(Bernstein et al. 2003), with the addition of an iodine cell behind
the spectrograph's entrance slit to model the instrumental profile and
to set an accurate reference wavelength scale (Butler et
al. 1996). The typical signal-to-noise of our spectra is about 130,
producing photon-limited uncertainties of 2 to 4 m s$^{-1}$. Two
additional sources of noise are present in the data. The first one is
the stellar jitter estimated in \S 2.  The second source of noise is
systematic instrumental errors, which for MIKE has a root mean square
(rms) deviation of 5 m s$^{-1}$, as derived from a subset of observed
stars that appear to have stable radial velocities over the time span
of the observations. A sample of stable stars measured with MIKE is
presented in Figures 1 and 2 of Minniti et al. (2008).

We obtained a total of 16 radial velocity measurements for HD 154672
and 24 measurements for HD 205739. Those measurements are summarized in
Tables 2 and 3, including the observation dates and the radial
velocity formal uncertainties introduced by photon-limited noise. The
data are represented in Figures 2 and 3.

For each data set, we modeled the radial velocities to fit single
planet Keplerian orbits using a Levenberg-Marquardt fitting
algorithm. In the case of HD205739 it was necessary to include an
additional variable linear trend to best reproduce the observed radial
velocity variations.
The parameter uncertainties of each best model fit were
estimated by running 1000 Monte Carlo trials on each data set, where
the model result of each trial was subtracted from the individual data
points and the residual velocities scrambled and added back
to the velocities predicted by the models, before running a new trial
fit. The adopted final uncertainties of each parameter are derived from 
the standard deviation of all the model trials.

The parameter values of the best Keplerian model fit for each target are
summarized in Table 4. The best model for HD154672 has an orbital
period of 163.94 $\pm$ 0.01 days, radial velocity semi-amplitude $K_{1}$ = 225
$\pm$ 2 m s$^{-1}$, and orbital eccentricity $e$ = 0.61 $\pm$ 0.03. The
$rms$ to the fit is 4.36 m s$^{-1}$, with a reduced 
$\sqrt{\chi^{2}_{\sigma}}$ = 1.60 relative to the radial velocity
formal uncertainties. 
Adopting a stellar mass of 1.06 $M_{\sun}$,
we derive a planetary mass of 4.96 $M_{Jup}$, and an average
relative separation for the system of 0.597 AU. The radial velocity data are
plotted in Figure~\ref{fig:154672}, together with the best Keplerian
model fit.

In the case of HD 205739, the best model has an orbital period of
279.8 $\pm$ 0.1 days, radial velocity semi-amplitude $K_{1}$ = 42
$\pm$ 3 m s$^{-1}$, and orbital eccentricity $e$ = 0.27 $\pm$
0.07. The radial velocities also show a substantial linear trend of
0.0649 $\pm$ 0.0002 m s$^{-1}$ per day that continues after over 3.5
years of observations. The $rms$ of the data to the fit is 8.67 m
s$^{-1}$, with a reduced $\sqrt{\chi^{2}_{\sigma}}$ = 2.13 relative to
the radial velocity formal uncertainties.  Adopting a stellar mass of
1.22 $M_{\sun}$, we derive a planetary mass of 1.37 $M_{Jup}$, and an
average semimajor axis for the system of 0.896 AU. The data with the
best Keplerian model fit are represented in
Figure~\ref{fig:205739}. As seen in the figure, the residuals to the fit
of a single planet plus a long term trend still appear large, showing
points that deviate about 2$\sigma$ from the average $\sim$4 m
s$^{-1}$ precision of the individual data points, however, an analysis
of the periodogram of those residuals reveals no significant peaks, so
we can not discard nor confirm the presence of additional shorter
period planets in this system with the current dataset.

The amplitude of the observed radial velocity variations for each star
is 10--100 times larger than the uncertainties of the individual
radial velocity measurements, what makes the possibility that the
detected signals are caused by noise fluctuations very unlikely.  We
quantitatively assert this statement by performing a false alarm
probability (FAP) analysis of the data using the method described by
\citet{Marcy05} (see \S 5.2), with the inclusion of possible linear
trends \citep{Wright07}. Figure~\ref{fig:FAP205739} shows the result
of 1000 FAP trial tests for HD205739. The FAP, i.e. the fraction of
trials of scrambled velocities that yields lower $\chi_{\sigma}$
than the best reported fit, is less than 0.1$\%$. A similar analysis of
the HD154672 data gives a negligible FAP ($\ll$ 1.0$\%$). In this last
case, the median $\chi_{\sigma}$ of the FAP histogram is about 30, the
first percently $\chi_{\sigma}$ is 16.1, and the minimum
$\chi_{\sigma}$ after 1000 trial tests is 12.5. None of the FAP trial
fits produces a $\chi_{\sigma}$ lower than the value reported above
for the best Keplerian fit for HD 154672.

\section{Discussion}\label{sec:discussion}

We present two new Jovian--mass planets orbiting metal rich stars.

HD 154672b is a fairly massive planet with a mass of M$\sin i$ = 4.96
$M_{Jup}$ and a very pronounced orbital eccentricity of $e$ = 0.61,
that causes the planet to move from 0.23 to 0.96 AU between periastron
and apastron.  The planet will therefore experience surface
temperature changes of about 300K along its orbit, reaching a maximum
temperature at periastron of about 600K, assuming the albedo of the
planet is low. If water is present in the atmosphere of HD 154672b,
it could transition between gaseous and liquid phases along the
planet's orbit.

When placed in the eccentricity vs. orbital period
parameter space diagram of known exoplanets illustrated in
Figure~\ref{fig:eccvsp}, HD 154672b shows an orbital eccentricity
larger than 90\% of the discovered planets, and is only the seventh
planet found with an orbital period shorter than 300 days and an
eccentricity larger than 0.6. Of the other six planets, four have been
found to be either in multiple-planet systems \citep[HD
74156;][]{Naef04, Bean08}, to have a brown dwarf companion \citep[HD
3651;][]{Mugrauer06}, to be part of a wide stellar binary system
\citep[HD 80606;][]{Naef01}, or to present a large radial velocity
trend induced by a distant body associated to that system \citep[HD
37605;][]{Wittenmyer07}. The high eccentricies in those cases can be
explained by either Kozai oscillations \citep{Kozai62} or chaotic
evolution of planetary orbits in multiple systems. Another planet in
this subgroup, HD 17156, has been recently reported to have a large
orbital axis misalignment with respect to the stellar rotation axis,
that can be best explained by gravitational interactions with other
planets \citep{Narita07}. There is however no evidence for additional
objets associated with the last planet in this subgroup, orbiting HD
89744.  The radial velocity curve of HD 154672 in
Figure~\ref{fig:154672} shows no trend nor significant residuals to
the fit that indicate the presence of other massive objects in that
system.

HD 205739b has a mass of M$\sin i$ =1.37 $M_{Jup}$, with an average
relative separation of 0.896 AU and an eccentricity of $e$ = 0.27. In
the eccentricity vs. orbital period diagram in Figure
~\ref{fig:eccvsp}, the parameters of this planet do not seem
atypical. For planets with orbital periods longer than 20 days, the
mean eccentricity is 0.29, therefore HD205739b has an orbital
eccentricity that is typical of detected planets that have not
experienced tidal circularization.  The separation of HD205739b from
its host star changes from 0.65 AU to 1.14 AU between periastron and
apastron. The maximum surface temperature of this planet is expected
to be of the order of 400K, and the amplitude of its surface
temperature change will only be of about 100K along the entire
orbit. One peculiarity of the radial velocity curve of HD 205739 is
the presence of a pronounced trend of 0.0649 m s$^{-1}$ per day,
indicating the presence of an additional outer body in the system with
an orbital period longer than the 3.5 year time span covered by our
observations, and a radial velocity semi-amplitude greater than 35 m
s$^{-1}$. Finally, the residuals of our best fit are a factor of two
larger than expected, what hints the possible presence of other bodies
in this system. However, more observations are needed to confirm this
hypothesis.

\acknowledgments We are thankful to the Chilean TAC for awarding
extended time to the Magellan N2K project.  MLM acknowledges support
provided by NASA through Hubble Fellowship grant HF-01210.01-A awarded
by the STScI, which is operated by the AURA, Inc., for NASA, under
contract NAS5-26555. MLM also acknowledges support from the Carnegie
Institution of Washington through a Carnegie Fellowship between
2004--2007, when most of the data presented in this work was
collected.  RPB has received support from NASA grants NAG5-12182 and
NNX07AR40G, NSF grant AST-0307668, and the Carnegie Institution of
Washington.  DAF acknowledges support from NASA NNG05B164G for the N2K
program and support from Research Corporation as a Cottrell Science
Fellow.  DM and PA are supported by CATA and FONDAP Center for
Astrophysics 15010003.  FCA has received support for this project from
NASA grant NAG5-12182.




\begin{table}[t]
\centering
\footnotesize
\caption{Stellar Parameters}
\label{tab:StPar} 
\begin{tabular}{ccc}
\hline\hline
Parameter& HD 154672 & HD 205739\\
\hline
V \dotfill&8.21&8.56\\
$M_{V} \dotfill$&4.11&3.78\\ 
B--V $\dotfill$&0.71&0.54\\ 
Spectral Type $\dotfill$&G3 IV &F7 V\\ 
Distance (pc) $\dotfill$& 65.8 & 90.3\\ 
$L_{bol}$/$L_{\sun}$ $\dotfill$& 1.88 & 2.3 \\ 
$[Fe/H] \dotfill$&0.26 (0.04)&0.19 (0.04)\\ 
$T_{eff} \dotfill$&5714 (30)&6176 (30)\\
$v\sin i$ $(kms^{-1})$ $\dotfill$& 0.54 (0.5)&4.48 (0.5)\\
$logg$ (cgs) $\dotfill$& 4.25 (0.08)&4.21 (0.08)\\
$M_{star}\ (M_{\sun})$\tablenotemark{a} $\dotfill$&(0.97) 1.06 (1.17)& (1.16) 1.22 (1.30)\\
$R_{star} (R_{\sun})$\tablenotemark{a} $\dotfill$&(1.18) 1.27 (1.37) &(1.23) 1.33 (1.43)\\
Age (Gyr)\tablenotemark{a} \dotfill& (6.92) 9.28 (11.44) & (1.72) 2.84 (3.76)\\
\hline\hline
\end{tabular}
\tablenotetext{a}{Values derived from evolutionary models.}
\end{table}

\begin{table}[t]
\centering
\footnotesize
\caption{Radial Velocities for HD 154672}
\label{tab:radvel1} 
\begin{tabular}{ccc}
\hline\hline
JD-2453000 & RV & $\sigma_{RV}$\\
(days) & (m$s^{-1}$)& (m$s^{-1}$)\\

\hline
   189.7132  &  -167.7  &  2.9 \\
   190.7083  &  -169.7  &  2.8 \\
   191.7204  &  -173.4  &  3.3 \\
   254.5062  &   149.1  &  2.6 \\
   596.6893  &   104.9  &  3.1 \\
   810.9097  &   -31.1  &  2.5 \\
   872.8136  &   234.8  &  2.5 \\
  1189.8750  &  -111.7  &  2.5 \\
  1190.8402  &   -83.3  &  2.8 \\
  1215.8605  &   216.6  &  2.5 \\
  1216.7893  &   217.8  &  2.6 \\
  1217.8725  &   224.9  &  2.8 \\
  1277.7025  &    41.6  &  2.7 \\
  1299.6210  &   -25.6  &  2.6 \\
  1339.5574  &  -172.5  &  4.1 \\
  1501.8960  &  -168.9  &  2.8 \\
\hline
\end{tabular}
\end{table}

\begin{table}[t]
\centering
\footnotesize
\caption{Radial Velocities for HD 205739}
\label{tab:radvel2} 
\begin{tabular}{ccc}
\hline\hline
JD-2453000 & RV & $\sigma_{RV}$\\
(days) & (m$s^{-1}$)& (m$s^{-1}$)\\

\hline
   189.8080  &   -33.3  &  4.0 \\
   190.8643  &   -27.7  &  4.4 \\
   191.8243  &   -31.7  &  4.0 \\
   254.6094  &   -77.0  &  3.6 \\
   550.8993  &   -39.0  &  4.3 \\
   551.8692  &   -57.0  &  3.9 \\
   655.6332  &   -27.5  &  3.2 \\
   657.5999  &   -10.4  &  3.6 \\
   658.6117  &   -11.9  &  3.4 \\
   685.5468  &    23.8  &  3.3 \\
   872.8832  &   -46.8  &  3.5 \\
   982.7455  &    38.4  &  4.0 \\
   988.7200  &    38.7  &  3.9 \\
  1013.6776  &    23.1  &  3.6 \\
  1066.5152  &   -13.4  &  3.7 \\
  1067.5231  &    -2.7  &  3.6 \\
  1216.9394  &    19.3  &  4.0 \\
  1277.8176  &    71.4  &  5.6 \\
  1299.8091  &    55.8  &  3.9 \\
  1300.7959  &    37.7  &  3.6 \\
  1338.7675  &    18.7  &  4.6 \\
  1339.7078  &    30.7  &  4.4 \\
  1397.5266  &    -2.0  &  4.7 \\
  1398.5065  &    -2.2  &  3.3 \\
\hline
\end{tabular}
\end{table}

\begin{table}[t]
\centering
\footnotesize
\caption{Orbital Parameters}
\label{tab:orbpar} 
\begin{tabular}{ccc}
\hline\hline
Parameter & HD 154672 & HD 205739\\
\hline
P (days) \dotfill&163.94 (0.01)  & 279.8 (0.1)\\
$T_{p}$ (JD+2450000.) $\dotfill$& 3045.3 (0.1)& 3390.7 (0.7)\\
$\omega (deg) \dotfill$& 265 (2)& 301 (8)\\
Eccentricity $\dotfill$&0.61 (0.03) & 0.27 (0.07)\\
$K_{1} (ms^{-1}) \dotfill$&225 (2) & 42 (3)\\
dv/dt (m$s^{-1}days^{-1}$) $\dotfill$& 0.0000& 0.0649 (0.0002)\\
$a_{rel}(AU) \dotfill$& (0.580) 0.597 (0.617) & (0.881) 0.896 (0.915)\\
$M\sin i$ ($M_{J}$) $\dotfill$ &(4.61) 4.96 (5.36) & (1.28) 1.37 (1.44)\\
$N_{obs} \dotfill$ & 16& 24\\
rms (m$s^{-1}$) $\dotfill$ &4.36& 8.67\\
Jitter (m$s^{-1}$) $\dotfill$ &4.0&4.5\\
reduced $\sqrt{\chi^{2}_{\sigma}} \dotfill$ & 1.60& 2.13\\

\hline\hline
\end{tabular}
\end{table}

\begin{figure}[hbt]
\epsscale{1.0}
\includegraphics[angle=90, width=6.5in]{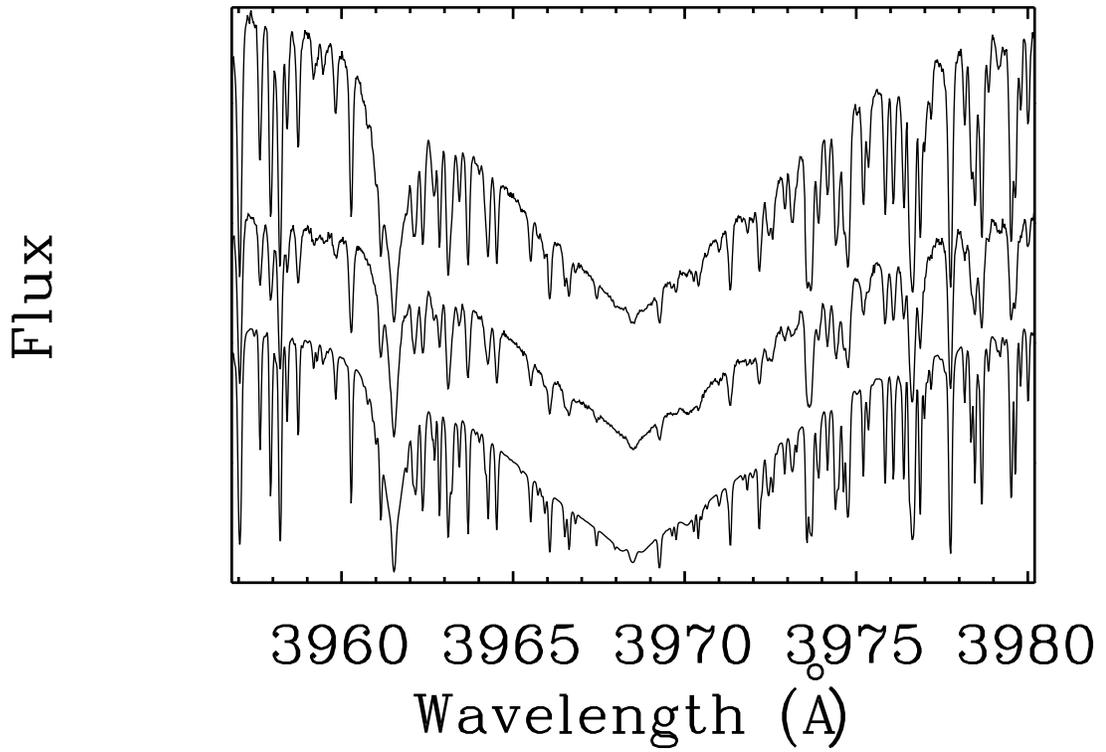}
\caption{CaII  H lines of HD 154672 (top) and HD 205739 (middle), compared to the same spectral line region for the Sun (bottom). The lack of emission in the line cores is consistent with low chromospheric activity levels.}
\label{fig:cak}
\end{figure}

\begin{figure}[hbt]
\epsscale{1.0}
\includegraphics[angle=90, width=6.5in]{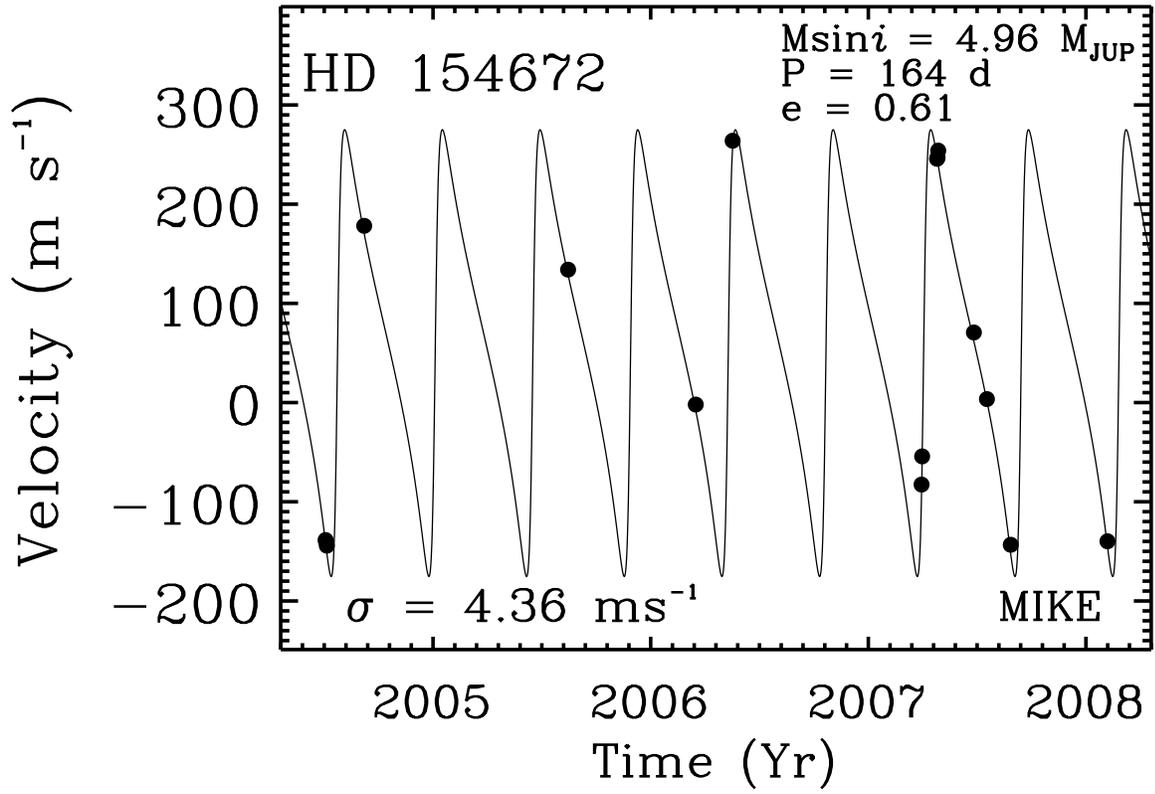}
\caption{Radial velocity measurements for HD154672 over 3.6 years. The minimum mass of the planet, assuming a stellar mass of 1.06 $M_{\sun}$, is $M \sin i$ = 4.96 $M_{Jup}$. There is no evidence for additional planets in this system.}
\label{fig:154672}
\end{figure}

\begin{figure}[hbt]
\epsscale{1.0}
\includegraphics[angle=90, width=6.5in]{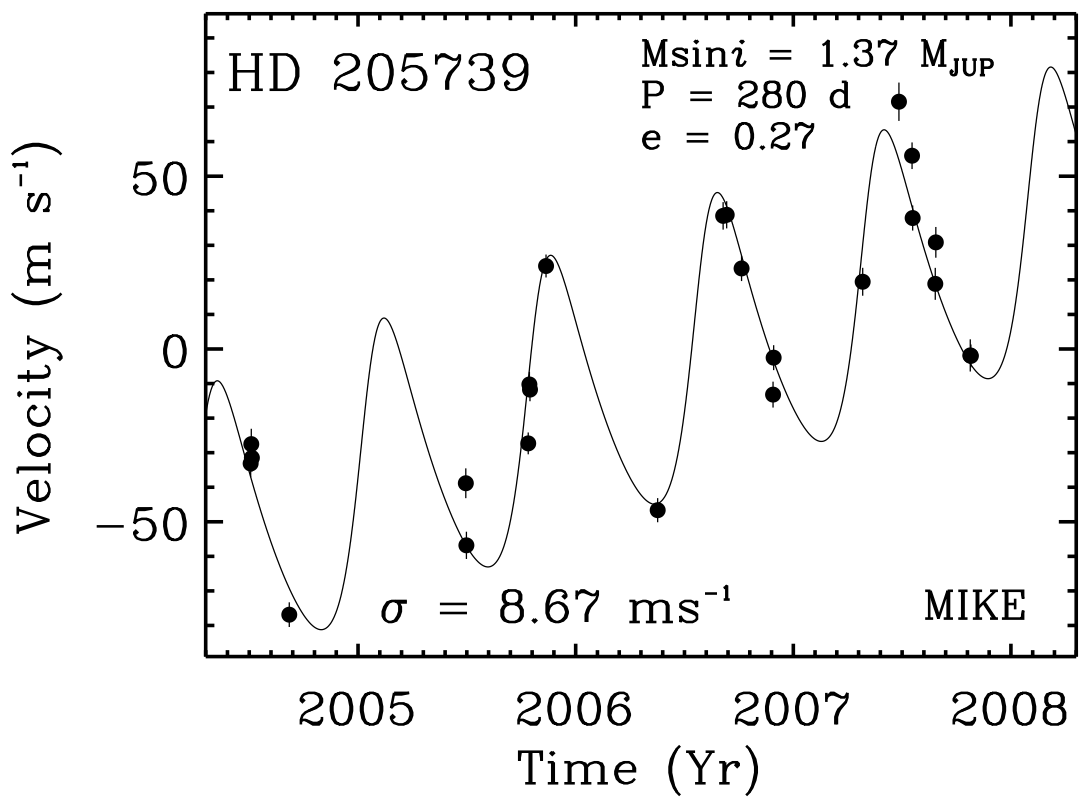}
\caption{Radial velocity measurements for HD205739 over 3.6 years. The minimum mass of the planet, assuming a stellar mass of 1.22 $M_{\sun}$, is $M \sin i$ = 1.37 $M_{Jup}$. The best Keplerian fit shows a significant trend of 0.0649 $\pm$ 0.0002 $ms^{1}$ per day, suggesting the presence of an additional outer body in the system.}
\label{fig:205739}
\end{figure}

\begin{figure}[hbt]
\epsscale{1.0}
\includegraphics[angle=0, width=6.0in]{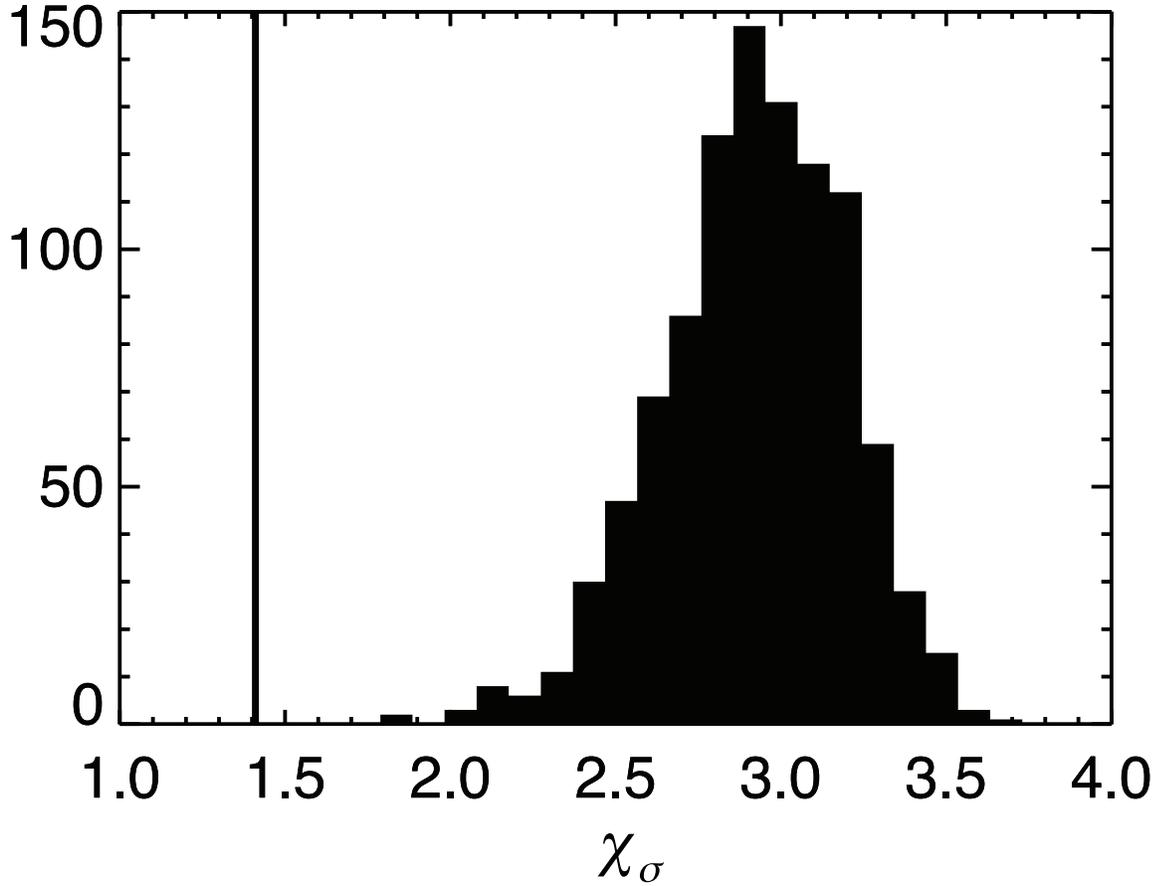}
\caption{Empirical evaluation of the FAP of the single-planet Keplerian fit plus a linear trend model for HD205739 reported in \S 3. The $\chi_{\sigma}$ of that fit is $\sim$ 1.4 and is indicated by the vertical line in the plot. The vertical axis shows the number of trials that produce a given $\chi_{\sigma}$. Less than 1 of the 1000 trial fits produce $\chi_{\sigma}$ values lower than the original time series of the observations, indicating that the FAP of the reported fit is less than 0.1$\%$.}
\label{fig:FAP205739}
\end{figure}

\begin{figure}[hbt]
\epsscale{1.0}
\includegraphics[angle=0, width=6.0in]{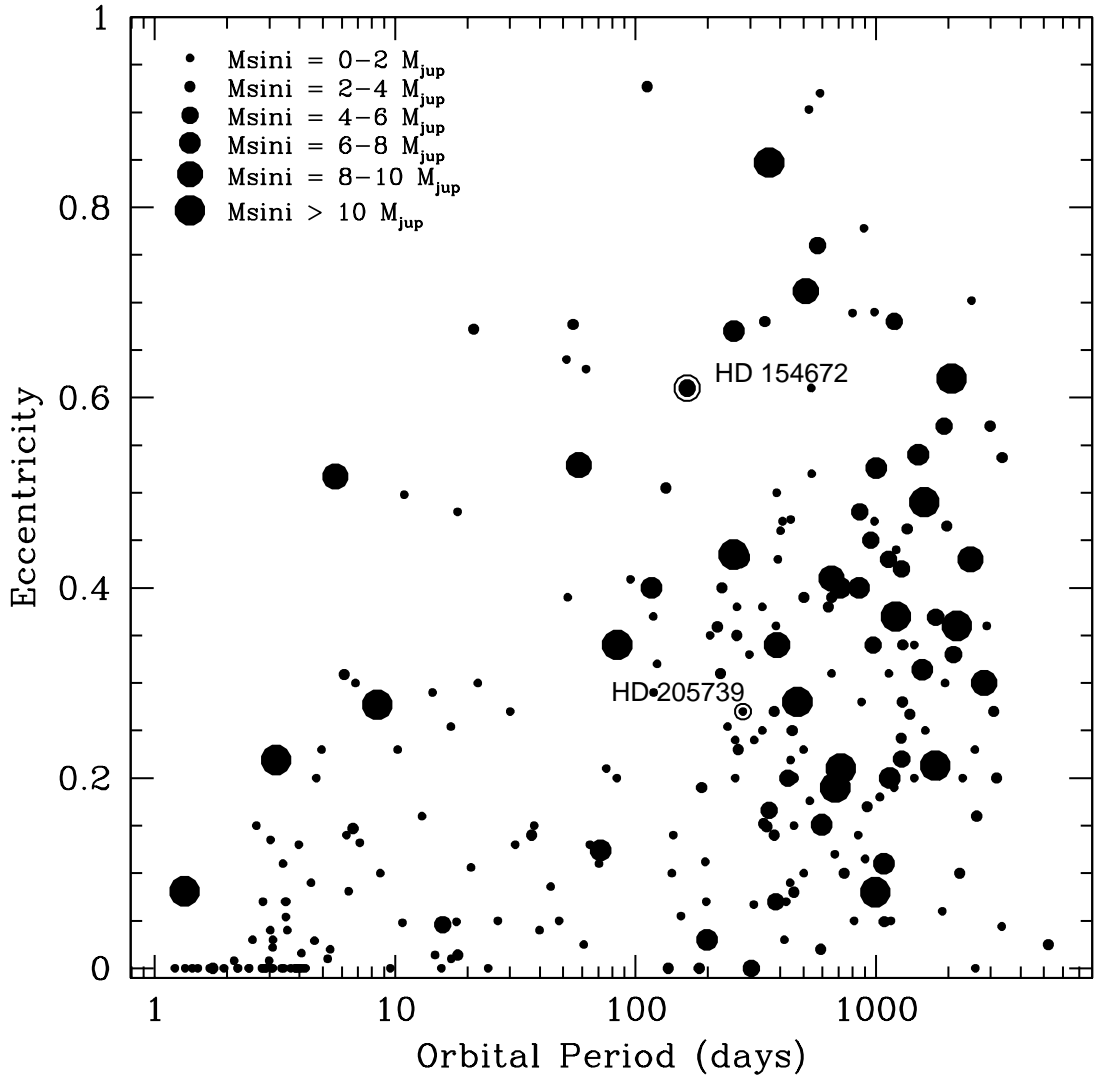}
\caption{Orbital eccentricity versus period diagram for known extrasolar planets. The size of the symbols scales with the planets' mass. The two open circles around the data symbols indicate the location of HD154672 and HD205739 in this diagram.
}
\label{fig:eccvsp}
\end{figure}

\end{document}